\documentclass[a4paper,12pt]{article}

\usepackage{graphicx}
\usepackage[hidelinks]{hyperref}
\usepackage{multirow}
\usepackage{amsmath}
\usepackage{booktabs}
\usepackage[right=20mm,left=20mm,top=20mm,bottom=30mm]{geometry}
\usepackage[width=.75\textwidth]{caption}

\newcommand{\footremember}[2]{%
	\footnote{#2}
	\newcounter{#1}
	\setcounter{#1}{\value{footnote}}%
}
\newcommand{\footrecall}[1]{%
	\footnotemark[\value{#1}]%
}

\newcommand{\orcidID}[1]{ \href{https://orcid.org/#1}{\includegraphics[scale=0.1]{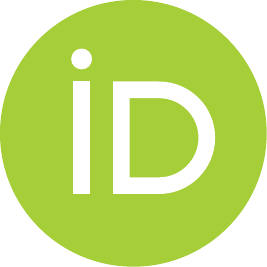}}}

\usepackage{indentfirst}

\begin{document}

\Large \textbf{Mastering 3D-detection of Extensive Air Showers in Cherenkov Light}

\vspace{\baselineskip}

\large Elena~A.~Bonvech\footremember{sinp}{Skobeltsyn Institute for Nuclear Physics Lomonosov Moscow State University, Moscow, Russian Federation}\orcidID{0000-0002-6878-357X},
Olga~V.~Cherkesova\footrecall{sinp}$^{,}$\footremember{fki}{Lomonosov Moscow State University, Space Research Department, Moscow, Russian Federation},
Dmitriy~V.~Chernov\footrecall{sinp}\orcidID{0000-0001-5093-9970},
Elena~L.~Entina\footrecall{sinp},
Vladimir~I.~Galkin\footrecall{sinp}$^{,}$\footremember{msu}{Lomonosov Moscow State University, Physics Department, Moscow, Russian Federation}\orcidID{0000-0002-2387-9156},
Vladimir~A.~Ivanov\footrecall{sinp}$^{,}$\footrecall{msu},
Timofey~A.~Kolodkin\footrecall{sinp}$^{,}$\footrecall{msu},
Natalia~O.~Ovcharenko\footrecall{sinp}$^{,}$\footrecall{msu},
Dmitriy~A.~Podgrudkov\footrecall{sinp}$^{,}$\footrecall{msu}\orcidID{0000-0002-0773-8185},
Tatiana~M.~Roganova\footrecall{sinp}\orcidID{0000-0002-6645-7543},
Maxim~D.~Ziva\footrecall{sinp}$^{,}$\footremember{vmk}{Lomonosov Moscow State University, Faculty of Computational Mathematics and Cybernetics, Moscow, Russian Federation}

\normalsize
%\author{Elena~A.~Bonvech\footremember{sinp}{Skobeltsyn Institute for Nuclear Physics Lomonosov Moscow State University, Moscow, Russian Federation}\orcidID{0000-0002-6878-357X}
%\and  Olga~V.~Cherkesova\footrecall{sinp}$^{,}$\footremember{fki}{Lomonosov Moscow State University, Space Research Department, Moscow, Russian Federation}
%\and Dmitriy~V.~Chernov\footrecall{sinp}\orcidID{0000-0001-5093-9970}
%\and Elena~L.~Entina\footrecall{sinp}
%\and Vladimir~I.~Galkin\footrecall{sinp}$^{,}$\footremember{msu}{Lomonosov Moscow State University, Physics Department, Moscow, Russian Federation}\orcidID{0000-0002-2387-9156}
%\and Vladimir~A.~Ivanov\footrecall{sinp}$^{,}$\footrecall{msu}
%\and Timofey~A.~Kolodkin\footrecall{sinp}$^{,}$\footrecall{msu}
%\and  Natalia~O.~Ovcharenko\footrecall{sinp}$^{,}$\footrecall{msu}
%\and Dmitriy~A.~Podgrudkov\footrecall{sinp}$^{,}$\footrecall{msu}\orcidID{0000-0002-0773-8185}
%\and Tatiana~M.~Roganova\footrecall{sinp}\orcidID{0000-0002-6645-7543}
%\and Maxim~D.~Ziva\footrecall{sinp}$^{,}$\footremember{vmk}{Lomonosov Moscow State University, Faculty of Computational Mathematics and Cybernetics, Moscow, Russian Federation}
%}

\title{Mastering 3D-detection of Extensive Air Showers in Cherenkov Light}

%\maketitle

\begin{abstract}

A new SPHERE seires complex extensive air showers detector is under development. The main goal of its mission is to study the mass composition of cosmic ray nuclei in the 1--100~PeV energy range at a new level. The already well-established telescope of Cherenkov light reflected from the snow-covered ice surface of Lake Baikal from an altitude of 500--1000 m will be supported by a detector of direct light pointed upward. Since the two detectors will study the same shower at different stages of its development, it could be called a 3D detection, which is completely new for the EAS method. The development is based on an extensive MC modeling of the shower and the detection process using the Supercomputer Complex of the Lomonosov Moscow State University.


\end{abstract}

\begin{center}
\footnotesize \textit{Submitted to the Supercomputing Frontiers and Innovations journal.}

\vspace{0.5\baselineskip}

\today
\end{center}

\section*{Introduction}
\label{sec:intro}
The history of cosmic ray (CR) study in general and the study of high-energy CR through the registration of extensive air showers (EAS), in particular, involves the continuous development of new detectors and approaches to experiment design. Since the first observations by Pierre Auger using simple Geiger counters and a coincidence scheme~\cite{Auger1938}, experiments have become more complex. The size of the detector arrays has increased, with the addition of new types of detectors dedicated to muon detection, leading to the development of the hybrid EAS detection technique. This approach has allowed more precise measurements of the CR energy spectrum and the first assessments of the mass composition of CR at high energies.

In addition to the registration of the charged particles of the EAS, the detection of optical components was proposed by A.E.~Chudakov (Cherenkov light~\cite{Chudakov1955} and fluorescent light~\cite{Chudakov1962}). He pioneered the use of Cherenkov light registration to find local gamma-ray sources~\cite{Zatsepin1961}, thus founding what would become gamma-ray astronomy.

The very first gamma-ray telescope (a set of four) had a single photoelectron multiplier and only registered an excess in the EAS count rate. However, the technique has evolved into modern projects~\cite{HESS,VERITAS,CTA} that analyze the properties of Cherenkov light images in telescope cameras~\cite{Hillas}, e.g. the angular distribution of Cherenkov photons. 

In 1974, A.E.~Chudakov proposed a new method of EAS registration --- detection of Cherenkov light reflected from the snow surface by an airborne detector~\cite{chu74:VKKL74}. The first successful realization of this method --- SPHERE-1~and~-2~\cite{Antonov2015a} --- registered the lateral distribution of Cherenkov photons with high accuracy. This distribution is also sensitive to the properties of the EAS primary particle~\cite{Chernov2017}. 

However, most of the above-mentioned registration and data analysis methods were based on analytical estimates or limited simulations. Later simulations became more complex but were still limited in power. Nowadays, with advances in computing power, data analysis can become more complex and more interrelated approaches to the experiment can be explored.

In the developing SPHERE-3 experiment we take a new approach to hybrid EAS registration: both spatial and angular Cherenkov light distributions, simultaneously at ground level and at altitude --- a 3D view of the shower. This task requires a wide use of available tools: parallel computing for large simulations and analysis of large data sets~\cite{Chernov2024}, neural networks for both data analysis~\cite{Bonvech2024} and detector optimization~\cite{Entina2024}.

\section{Experiment features resulting from its main goal}

In general, the SPHERE-3 detector inherits its main features and scientific goals from its predecessor, the SPHERE-2 detector. However, we are shifting the focus of our research towards the primary cosmic ray composition, which will be the primary goal of the new experiment. The new 3D approach can help us greatly in this task.

The study of the mass composition of ultrahigh-energy primary cosmic rays is a challenging problemto be addressed in an EAS registration experiment. In order to obtain data on the mass of the primary nucleus, we need to find a measurable quantity that depends on the mass but is virtually independent of hadron interaction models which, remain a major obstacle to solving the cosmic ray mass composition problem.

Our experience with EAS Cherenkov light detection allows us to formulate some general requirements for such quantities. First, the quantity should be related to the shape of the observable distribution rather than to the absolute value. Second, the quantity must be a combination of integrals over parts of the measured image, or alternatively a parameter of an approximation to the image, in order to be less sensitive to image fluctuations. These two requirements are met by any measurable EAS Cherenkov light (CL) characteristic. As a result, the mass sensitive parameter based on the EAS CL is only slightly dependent on the interaction model.

\begin{figure}[bth]
    \centering
    \includegraphics[width=.7\textwidth]{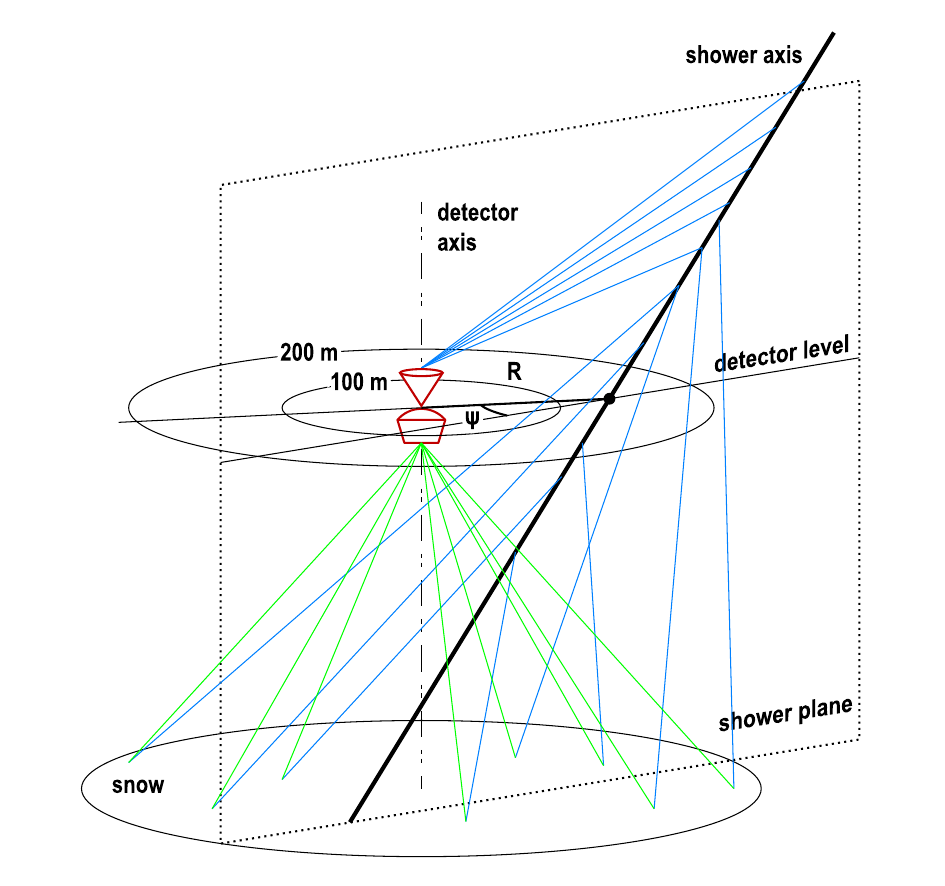}
    \caption{EAS detection at two levels. Reflected light (green) is collected by the lower detector of the SPHERE-3, direct light (blue) is accepted by the upper one.}
    \label{fig:3D_detection}
\end{figure}

Considering the EAS simulation results, we decided to use the image from the reflected light telescope to estimate $E_0$, the axis direction, the core location on the snow and the primary mass (see Fig.~\ref{fig:3D_detection}). The image from the direct light detector is used for a more accurate evaluation of the axis direction and the primary mass~\cite{Galkin2024}. The size of the long dimension of the light spot is a tentative mass parameter, but a better one is being sought. However, the sensitivity of the spot length to mass is greater than that of the image steepness in the telescope, forcing us to modify the measurement strategy to maximize the fraction of events with images in both detectors. At an altitude of 500~m, the fraction is about 30\% and decreases with altitude, making 500~m the preferred altitude for SPHERE-3 flights.

\section{Construction blocks of the experimental data handling procedure}
The entire shower parameters evaluating process involves a series of specialized algorithms that estimate a particular parameter based on data from a single detector. These algorithms are then grouped according to their data source.

\subsection{Based on the reflected light telescope images}

\subsubsection{Axis position on the snow and axis direction evaluation}
The arrival direction of the primary particle is determined from the temporal characteristics of the EAS CL registered in pixels of the telescope mosaic. A Cherenkov pulse is identified in each pixel of the mosaic and its maximum is located. The corresponding time is related to the time delay of the shower at the corresponding location on the snow, taking into account the position of the pixel in the mosaic and the altitude. The time taken for the photon to travel from the observed snow spot to the telescope is also taken into account.

This process results in a set of time delays at different points on the snow surface forming a EAS CL temporal front. This front is approximated by a quadratic polynomial in the EAS coordinate system, creating a paraboloid of rotation around the shower axis. This method allows us to estimate the primary particle arrival direction with an accuracy of about 1--2$^\circ$.

The shower axis location on the snow is determined by the spatial distribution of the CL photons on the telescope mosaic. For each pixel, the integral of the Cherenkov pulse is estimated. The pixel with the highest integral is then identified. If this maximum is close to a true maximum of the EAS CL distribution (and not caused by fluctuations, as discussed in the next paragraph), the shower axis is defined as the center of mass of two rows of pixels around the maximum pixel. If the maximum pixel is close to the edge of the mosaic, the center of mass is calculated using only one row. If the maximum pixel is at the edge of the mosaic, the event should be ignored as the shower axis cannot be reliably determined as may be outside the field of view of the detector.

Once the shower axis has been determined, its coordinates are mapped onto the snow surface and expressed in the telescope's coordinate system (relative to the telescope's vertical axis). 

\subsubsection{Primary energy evaluation procedure}
The energy evaluation is based on the integral $Q$ of the approximation function, which depends on the distance $R$ from the detector axis to the shower axis on the snow. The set of dependencies $Q(R, E_0)$ is constructed by approximating the artificial image data. The dependency $E_0(R, Q)$ is constructed as an inverse interpolation of $Q(R,E_0)$. The quantities $R_\text{exp}$ and $Q_\text{exp}$ are obtained from the approximation of an image to be processed. By substituting $R\text{exp}$ and $Q_\text{exp}$ into the formula, we obtain a primary energy estimate $E_0^\text{est}$:

\begin{equation}
     E_0^\text{est}= E_0( R_\text{exp} , Q_\text{exp} ).
\end{equation}

To accurately determine the spatial distribution function (SDF), it is important that the distribution axis is within the mosaic and aligned with the image axis. Otherwise, errors may occur in the determination of $Q_\text{exp}$ and $R_\text{exp}$. However, it is not always possible to discard an unsuitable image. If the SDF axis is outside the mosaic, the shape of the image may increase towards the boundary in the direction of the SDF axis. This can cause the maximum in the image to be on the mosaic boundary and the approximation to show that the SDF axis is outside the field of view. However, due to fluctuations in the number of photons comparable to the number of photons in the segment, the maximum of the image can shift from its true position. Therefore, during approximation, this local maximum can be mistaken for the SDF axis (we will refer to such maximums and axes as false ones). This leads to an incorrect distribution shape and therefore the value of $Q_\text{exp}$ and consequently the energy will be underestimated.

The method for determining false maxima is developed on the basis of model data. It is based on finding the boundary of the distribution of the value characterizing the image for true and false maxima. The boundary is chosen to reject as few images as possible with a correctly defined axis, and as many as possible with a false axis. To make the selection criterion universal, relative features that are weakly dependent on the primary parameters are considered. If the criterion is still dependent on the primary energy, its relation to a measurable quantity correlated with the energy is established and used to adjust the criterion.

Several quantities have been considered. At the moment, the quantity chosen for the method is:

\begin{equation}
    L=\left(\frac{f_\text{surf}}{q_1}-\frac{f}{q_2}\right)\times100,
\end{equation}

\noindent where $f_\text{surf}$ and $f$ are the values of the function characterizing the accuracy of the plane approximation and the axially symmetric function, $q_1$ and $q_2$ are the number of degrees of freedom. The separation boundary for this value $L$ is linearly dependent on $f$.

In the case of a sample of 200 events, with 100 false and 100 true maxima, the average energy error is 28\%. After applying the false maximum detection method, the average error is resudecd to 17\%. The method eliminates 73\% of the false maxima with a 3\% error in eliminating true maxima.

\subsubsection{Primary mass estimation procedure}
It is well known that the depth of the maximum of a EAS depends directly on the mass number $A$ of the primary particle:
\begin{equation}
 \Delta{}X_\text{max}\approx-D\ln A,
\end{equation}
which we simplify to $\Delta{}X_\text{max}\approx-\ln A$.

This means that the pattern of Cherenkov light on the ground (and in the detector) will differ as a function of the mass $A$, all else being equal. For lighter nuclei (such as protons) we expect a wider pool of light with a less pronounced peak, whereas for heavier nuclei (such as iron) we expect a narrower pool with a more pronounced peak.

Using a lateral distribution function (LDF) $F(R)$, which approximates the shape of the image in the focal plane of the telescope, we can quantify these differences and determine the primary particle type. We consider three types of nuclei: protons (p), nitrogen (N) and iron (Fe).

Given the LDF $F$, we introduce a one-dimensional mass sensitive criterion:

\begin{equation}
    C=\frac{\int\limits_{0}^{r_1}F\, \mathrm{d}R}{\int\limits_{r_1}^{r_2}F \, \mathrm{d}R},
\end{equation}

\noindent where $r_1$ and $r_2$ are the radii of two concentric circles centered on the intensity peak with $r_2\geq r_1$.

By choosing appropriate values for $r_1$ and $r_2$ for each event, we can formally measure the width and height of the intensity distribution of the shower image on the camera. In our current implementation, we scan through the ranges $r_1 \in \left[10,110\right]$ and $r_2 \in \left[110,270\right]$.

The optimal ($r_1$, $r_2$) pair is determined by minimizing the type II error in two binary classifications: p vs. N and N vs. Fe. Once these radii have been determined, we construct the distribution of $C$-values for simulated events and use it to define separation boundaries in a training sample. These boundaries are then used to classify the primary particle in each measured EAS image.

\subsection{Based on the direct light detector images}

\subsubsection{Primary direction estimation using the angular distribution of Cherenkov light}
\label{sec:direct_direction}
\begin{figure}[th]
    \centering
    \includegraphics[width=90mm]{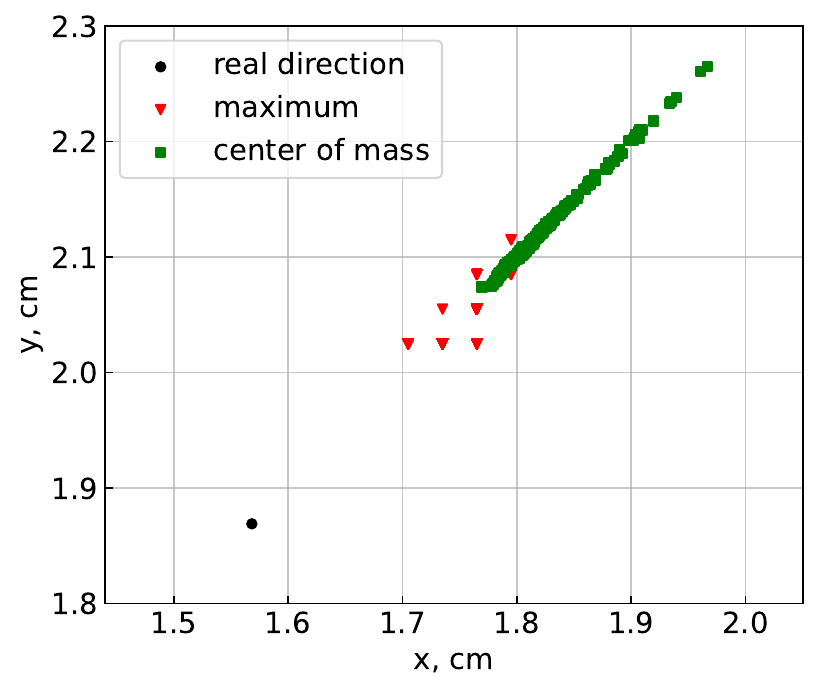}
    \caption{The distribution of maxima and centers of mass of the images for a distance $R=100$~m from the shower axis to the detector and a detector azimuth $\psi=45^{\circ}$. Flight altitude 500 m above the snow level.}
    \label{fig:centers_distrib}
\end{figure}

The primary particle direction is determined from the characteristics of the Cherenkov image captured on the sensor after the Cherenkov photons have passed through a 400~cm$^2$ lens with a focal length of 64~cm. Currently, the primary particle arrival direction is evaluated using a key point, which is the center of mass or the location of the maximum intensity of the light spot. The distribution of intensity maxima and centers of mass of the images for a given distance $R$ from the shower axis to the detector and a given detector azimuth $\psi$ is shown in Fig.~\ref{fig:centers_distrib}.

In determining the primary particle arrival direction, a systematic shift occurs as a function of $R$. This shift can be estimated as the average difference between the actual direction and the key point direction. A comparison of the uncertainties in the primary particle arrival directions before and after eliminating the shift for different $R$ is given in Tab.~\ref{tab:direction_uncert}.

\begin{table}[th]
    \centering
    \caption{The uncertainties of the primary particle direction estimates before and after the shift elimination for different $R$.}
    \label{tab:direction_uncert}
    \begin{tabular}{ccccccc}
        \toprule
        \multirow{2}{*}{$R$, m} & \phantom{000} & \multicolumn{2}{c}{By maximum} & & \multicolumn{2}{c}{By center of mass} \\
        \cmidrule{3-4}\cmidrule{6-7}
        & & before & after & & before & after \\
        \midrule
         100 m & & 1.28 & 0.10 &  &2.28 & 0.22 \\
         140 m & & 1.46 & 0.20 &  &2.78 & 0.32 \\
        \bottomrule
    \end{tabular}
\end{table}

\subsubsection{Primary mass evaluation by the angular distribution of Cherenkov light}
\label{sec:direct_mass}
One of the challenges of the SPHERE-3 project is to find an optimal criterion for classifying primary nuclei. Currently, they are divided into three groups based on the length of the image: proton, nitrogen and iron. 

The classification procedure for the p-N pair implies that an event is caused by a proton if the spot length exceeds the critical value, and by a nitrogen nucleus if the opposite is true. For the N-Fe pair, an event with a spot length above the critical value is attributed to N, while an event with a shorter spot length is attributed to Fe.

For a fixed primary energy, the spot length depends on several parameters: the distance from the detector to the shower axis at flight level, the azimuthal angle of the detector with respect to the shower plane, and the absolute threshold. The latter is the number of photoelectrons per histogram bin, so that bins with contents above this threshold are taken into account. We have shown (see Tab.~\ref{tab:direct_mass_calssification}) that using a common criterion for all distances and azimuths can lead to misclassification errors exceeding 0.3.

\begin{table}[tb]
    \centering
    \caption{Probabilities of misclassification of the primary particles using different criteria (see text). Here p is the probability of proton misclassification, p-N denotes probability to take N for proton, N-Fe --- probability to take N for Fe, Fe --- probability of iron nuclei misclassification.}
    \label{tab:direct_mass_calssification}
    \begin{tabular}{lccccc}
        \toprule
        Approach & & p & p-N & N-Fe & Fe  \\
        \midrule
        common criterion & & 0.44 & 0.14 & 0.42 & 0.11 \\
        system of criteria & & 0.34 & 0.23 & 0.32 & 0.16 \\
        absolute threshold dependent on $R$ & & 0.25 & 0.26 & 0.23 & 0.24 \\
        with double detection method & &0.39 & 0.25 & 0.39 & 0.26 \\        
        \bottomrule
    \end{tabular}
\end{table}

\begin{figure}[ht]
    \centering
    \includegraphics[width=.7\textwidth]{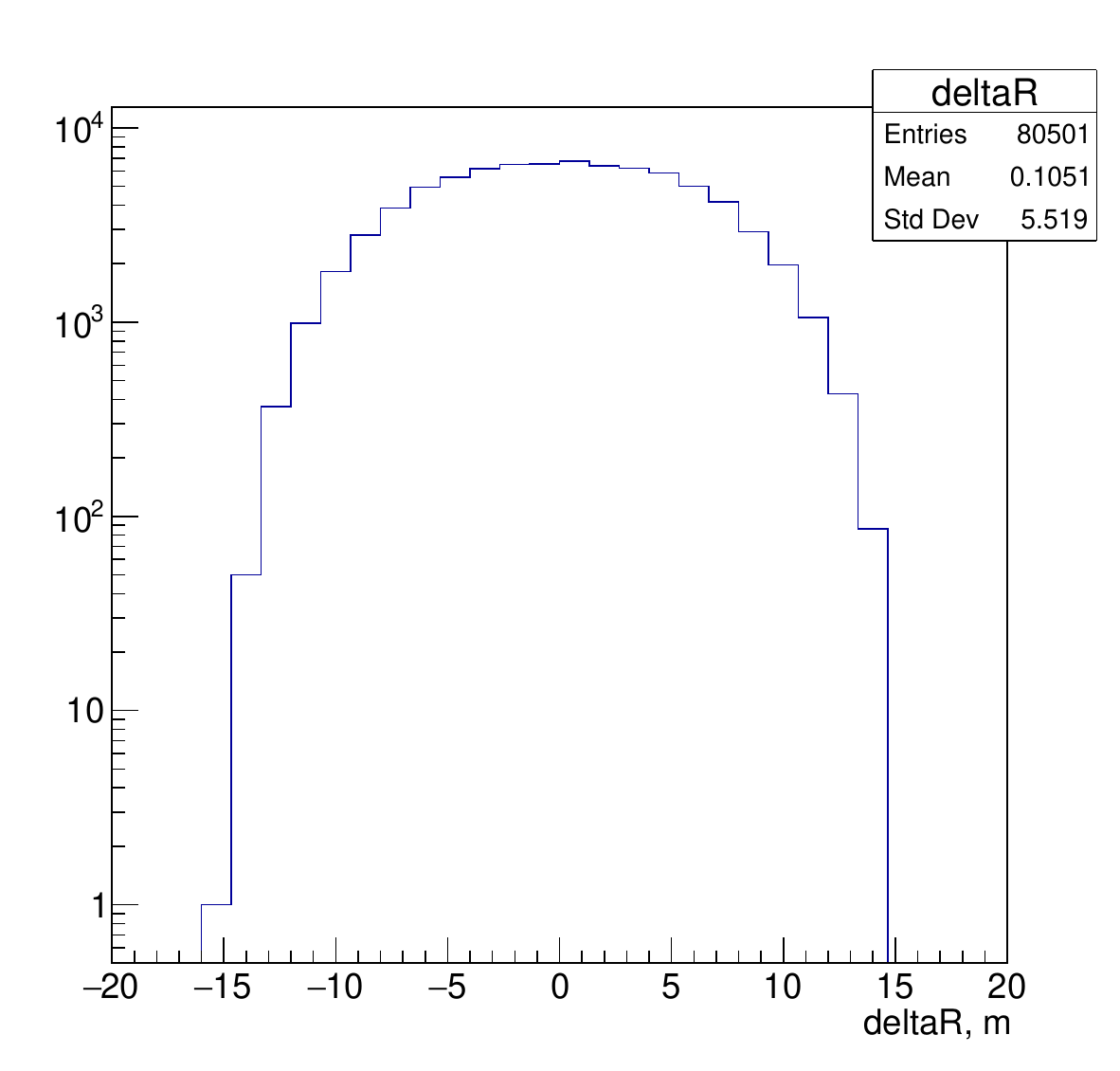}
    \caption{The distribution of the uncertainties of the distances $R$ from the shower axis to the detector at the altitude of 500~m.}
    \label{fig:uncert_distrib}
\end{figure}

It was therefore decided to develop a system of criteria based on the critical Cherenkov image length, which depends on the distance $R$ from the detector to the shower axis and the azimuth $\psi$ of the detector. The system considers distances $R$ in the range 100--200~m and azimuths $\psi$ in the range 0--360$^\circ$. The $R$-range is divided into five bins and the $\psi$-range into 24 bins. The number of $R$ bins was chosen because of the uncertainty in $R$ the flight level (see Fig.~\ref{fig:uncert_distrib}).

The size of the azimuthal bins is 15$^\circ$, which allows the corresponding criterion to have approximately the same misclassification error as one with smaller bins.

The use of this system reduces the probability of misclassifying primary particles, as shown in Tab.~\ref{tab:direct_mass_calssification}.

Since the number of photons in the image at fixed primary particle energy $E_0$ depends strongly on the distance from the detector to the shower core $R$ at the detector altitude, it is possible that using an absolute threshold that varies with $R$ can reduce the chance of misclassifying the primary particle. We found that applying a higher threshold for $R<150$~m and a lower threshold for $R>150$~m actually reduced the classification errors to the levels shown in Tab.~\ref{tab:direct_mass_calssification}.

\section{Joint criterion of primary mass}
\label{sec:joint_cri}
Now we come to the point where the power of the new development needs to be demonstrated. The ability of the reflected light telescope to distinguish between different nuclei is limited. The direct light detector can analyze the Cherenkov angular image of the EAS with better results for axis direction and mass classification. However, it is still not reasonable to ignore even a small assistance in such a sensitive matter as the primary mass study. Therefore, a joint criterion has been constructed based on the two criteria already described.

The training and testing of the criteria requires the generation of appropriate samples that reproduce the situation of double detection. Specifically, a shower must be visible to both the reflected light telescope and the direct light detector. At an altitude of 500~m, this means that the shower axis must hit the snow at a maximum distance of 230~m from the detector axis. The shower axis at 500~m altitude must cross the ring (100~m, 200~m) around the detector. Close passes produce images that are too compact and difficult to classify. Distant showers produce faint images with fluctuating dimensions and shapes.

For each EAS event, a number of clones with different core locations on the snow are created within 230~m radius circle centered on the detector axis. If the axis of a clone intersects the ring at the flight altitude the clone is accepted and its images in both detectors are calculated and classified using the appropriate criteria. Otherwise the clone is rejected.

\begin{figure}[ht]
    \centering
    \includegraphics[width=.7\textwidth]{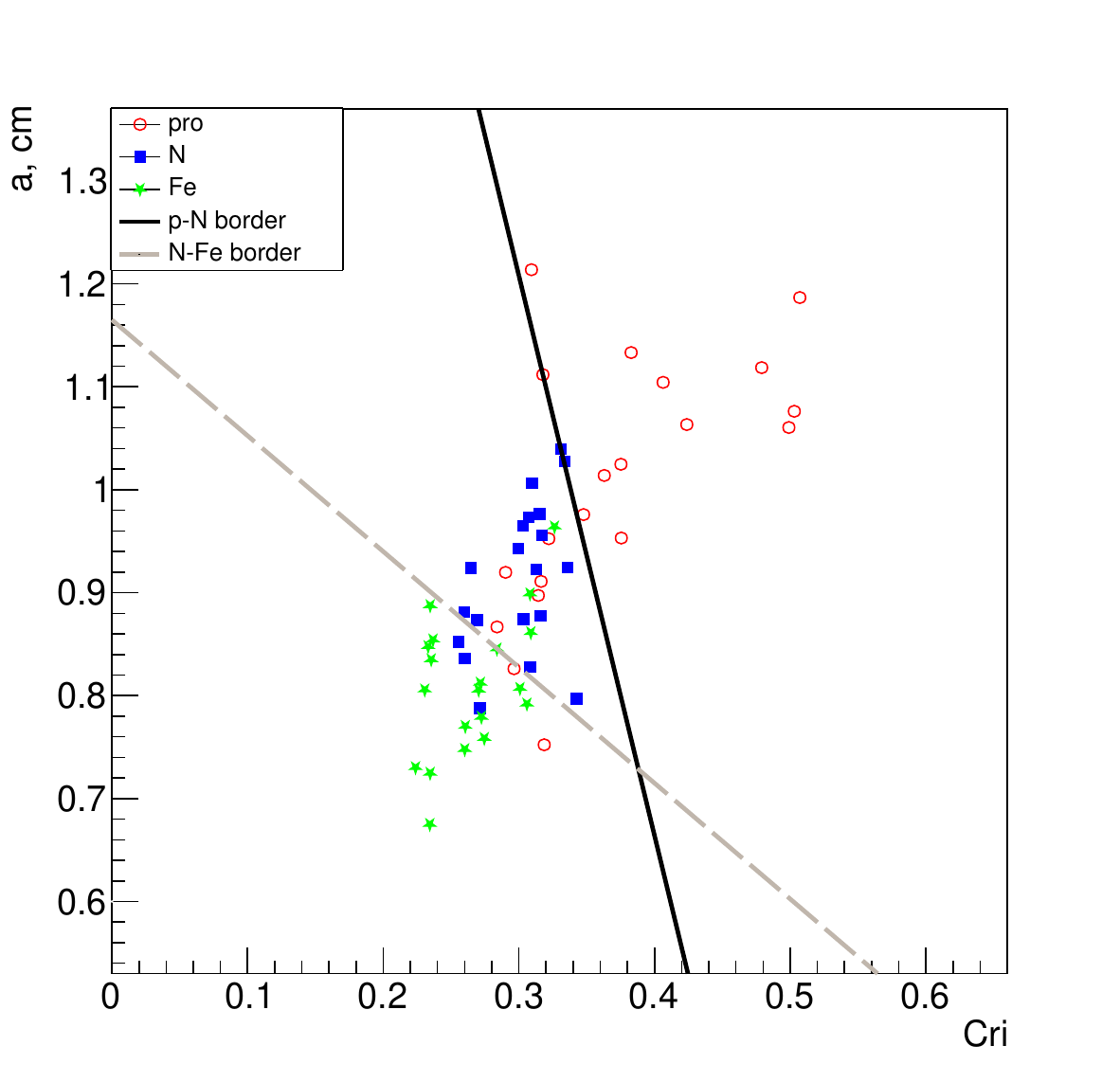}
    \caption{Straight lines separating the pairs p--N and N--Fe during dual detection. Here $a$ is the feature made of the direct light image, $Cri$ is the feature of the reflected light image. Detector altitude of 500~m above the snow level, distance range 100--112.5~m, azimuthal range $0^\circ$--$15^\circ$.}
    \label{fig:direct_mass_cri}
\end{figure}

For each clone, the values of both features are combined to form a vector. This creates a point in a two-dimensional feature space, which can then be used to search for the optimal linear decision function $ax+b$. The line that gives the lowest probability of misclassification is considered optimal. The system also uses the criteria of the direct light detector with the absolute threshold depending on the distance $R$. This creates a system where the choice of linear decision function is determined by the $R$ value at altitude and detector azimuth. Fig.~\ref{fig:direct_mass_cri} shows two decision functions from the system.

The dual detection method currently produces the primary particle misclassification probabilities shown in Tab.~\ref{tab:direct_mass_calssification}.

As can be seen from the values, the probability of misclassification is quite high. In order to reduce them, it is planned to modify the function to be minimized and to improve the gradient descent algorithm.

\section{Self-consistent procedure for EAS primary parameter assessment}
	 	 	 	 	 	
Images of an EAS in the reflected Cherenkov light telescope and the direct light detector contain a vast amount of data on the primary parameters of the shower, such as the energy $E_0$, the arrival direction $(\theta,\phi)$ and the mass $A$ of the nucleus that initiated the shower. 

\begin{figure}[ht]
    \centering
    \includegraphics[width=.7\textwidth]{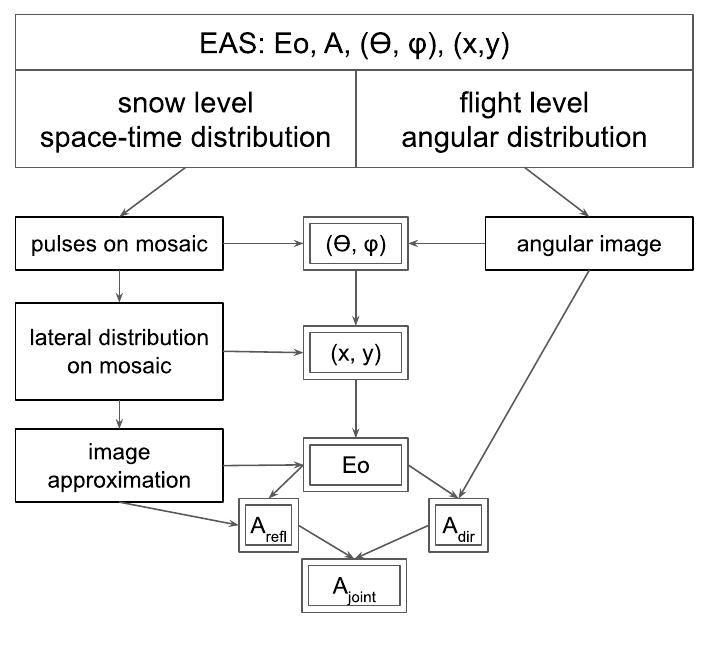}
    \caption{Scheme of the self-consistent procedure to estimate  the primary parameters. Some parts of it will be looped. At the final stage one might require to iterate the whole procedure.}
    \label{fig:est_scheme}
\end{figure}

We have already developed specific methods to determine $E_0$, $(\theta,\phi)$ and axis location separately using data from the telescope, without considering $A$. We have also developed a method to determine $(\theta,\phi)$ from the direct light image, ignoring $E_0$ and $A$. For both the telescope and the detector, we have developed procedures to estimate $A$ using only a rough estimate of $E_0$. Now we need to develop a comprehensive procedure that determines all the parameters consistently (see Fig.~\ref{fig:est_scheme}).

Inevitably, one must start with estimates of $E_0$, $(\theta,\phi)$ and the axis location based on the image in the telescope, even if the image from the upper detector is available. First, the image of the reflected light must be corrected for the optical distortions in the telescope and approximated by an axially symmetric function. Typically, the image maximum corresponds to the shower core region on the snow, and so does the approximation maximum. Events with axes outside the field of view of the telescope may mimic the true maximum due to fluctuations. To eliminate these cases, certain measures are taken: 1) images with maxima at the edge of the mosaic are discarded, and 2) in addition to the approximation using aforementioned function, another approximation of the image is made using a plane, and the results of the two fits are compared, which helps to filter out false maxima.

After confirming that the image maximum is real, the location of the shower core on the snow is calculated. A fit to the time delay of the light contributions to each pixel gives an estimate of the shower arrival direction with some accuracy.

A rough estimate of $E_0$ (ignoring the $A$ estimate) is obtained by integrating the true approximation over a given circle. Finally, a ratio of two integrals of the approximation function over a central circle and an outer ring is calculated, which represents the steepness of the approximation and is sensitive to the primary particle mass.
 	 	 	 	 	
If the direct light image is not available, the final step in the process is an estimation loop: the direction and mass estimates are used to refine the energy estimate; the mass estimate is recalculated taking into account the new energy estimate, and so on. The axis position on the snow remains unchanged.

If the direct light image is suitable for analysis, the axis of the shower is extrapolated to altitude. Its coordinates $(R,\psi)$ are then estimated, which helps more accurately determine the shower axis direction. The shower axis position and direction are then adjusted in a recursive, self-consistent manner at the flight height.

The primary energy estimate is also important (see section~\ref{sec:direct_direction}). Therefore, one must repeat the direction evaluation with a given energy estimate and the energy evaluation with a given direction estimate until both estimates converge.

Until now, the energy estimate has not been affected by the mass estimate. To account for mass, a joint procedure for mass evaluation must be applied, based on mass sensitive parameters of both the reflected light telescope image and the direct light detector image.

The axis position is important for the primary mass estimation, as explained in the previous section~\ref{sec:direct_mass}. The uncertainty in the axis position, ensured by the self-consistent recursive procedure, amounts to approximately 10~m for 500~m altitude, which is sufficient for the complex mass criterion system to divide the detected events into three groups by mass and presumably even for the construction of a mass regression with more sensitive parameters, which have yet to be found. The mass criterion to be used is chosen from a set of $5 \times 24 = 120$ different ones, depending on the position of the shower relative to the detector.

The next step is to combine the mass criterion with the reflected light criterion into a joint one, as described in section~\ref{sec:joint_cri}. The mass estimate should now be used to correct the energy estimate. This correction may affect the direction and mass estimates, so the whole procedure must to be repeated.

Perhaps we will eventually be able to combine all of the above criteria and relations into a single function. For now, the procedure described above seems to make some sense, especially since we have already fine-tuned some of its parts.

\section{Conclusion}

The design of the new SPHERE-3 complex detector is being optimized to meet the stated goal of studying the mass composition of primary cosmic rays in the 1--100 PeV energy range. This optimization is only possible through a large number of Monte Carlo simulations of the EAS and the processes inside the detector.

We have already explored several detector designs and experimental strategies, and now present some preliminary conclusions on how to preceed.

A special feature of the SPHERE-3 detector is that the telescope for the registration of the reflected Cherenkov light will be supported by a detectorforf the direct light pointed at the zenith and capable of analyzing the Cherenkov images in detail. The strategy behind this is that a certain fraction of the detected events will contribute images to both detectors, providing an unprecedented level of information about EAS at two different depths in the atmosphere. 

We have shown the importance of the additional information provided by the direct light detector, and this has led us to choose an optical scheme and experimental design that will maximize the fraction of events seen by both detectors.

\section*{Acknowledgements}
The research was carried out using the equipment of the shared research facilities of HPC computing resources at Lomonosov Moscow State University~\cite{Supercomputer}.

This work is supported by the Russian Science Foundation under Grant No. 23-72-00006, https://rscf.ru/project/23-72-00006/

\bibliography{references.bib}

\end{document}